\documentclass[aps,prl,twocolumn,showpacs,preprintnumbers,superscriptaddress,amsmath,amssymb]{revtex4}
\usepackage{graphicx}
\usepackage{dcolumn}
\usepackage{bm}
\usepackage{ulem}
\usepackage{color}

\newcommand{\s}{{\sigma}}

\def\be{\begin{eqnarray}}
\def\ee{\end{eqnarray}}

\newcommand{\nn}{\nonumber\\}

\newcommand{\Eq}[1]{Eq.~(\ref{#1})}

\newcommand{\p}{\partial}

\newcommand{\ra}{\rightarrow}

\newcommand{\Fig}[1]{Fig.~(\ref{#1})}

\begin{document}

\title{The effects of interaction on quantum spin Hall insulators}
\author{Dung-Hai Lee}
\affiliation{
Department of Physics,University of California at Berkeley,
Berkeley, CA 94720, USA}
\affiliation{Materials Sciences Division,
Lawrence Berkeley National Laboratory, Berkeley, CA 94720, USA}

\date{\today}

\begin{abstract}
We study the $S_z$-conserving quantum spin Hall insulator in the presence of Hubbard $U$ from a field theory point view. The main findings are the following. (1) For arbitrarily small U the edges possess power-law correlated antiferromagnetic XY local moments. Gapless charge excitations arise from the Goldstone-Wilczek mechanism. (2) The decaying exponent of the XY correlation is $\ge 2$ hence normally space-time vortices should proliferate. (3) For thermodynamic samples, vortex proliferation is prohibited by the conservation of edge charge. (4) For a sample with finite width electron tunneling between opposite edges allows vortex proliferation hence is a strongly relevant perturbation (more relevant than in the non-interacting theory). (5) The phase transition between the antiferromagnetic and the topological insulating phases is triggered by the condensation of magnetic excitons. 
\end{abstract}

 \maketitle

The subject of topological insulator (TI) has attracted considerable attention recently\cite{moore}.
A signature of this type of band insulator is the presence of itinerant boundary states in the bulk band gap. Moreover, unlike those in usual band insulators, these itinerant in-gap states are robust against any modification of the free-electron Hamiltonian so long as they (i) respect time reversal symmetry, and  (ii) do not close the bulk bandgap. For example, three dimensional TIs possess in-gap surface states whose dispersion consists of an odd number of Dirac cones\cite{topo,ARPES}. Because of the robustness, these surfaces states can evade Anderson's localization in the presence of (time-reversal invariant) disorder\cite{ryu}. As another example, two dimensional TIs (or the quantum spin Hall insulator (QSHI)) possess odd pairs of time-reversal conjugate, counter-propagating, edge states\cite{km1,km2,zb,konig} - the ``helical edge modes''.  At the present time non-interacting TIs are fairly well understood. What remains open is the effect of electron-electron interaction on TIs\cite{int}.

\begin{figure}[tbp]
\begin{center}
\vspace{-0.4 in}
\includegraphics[scale=0.38]
{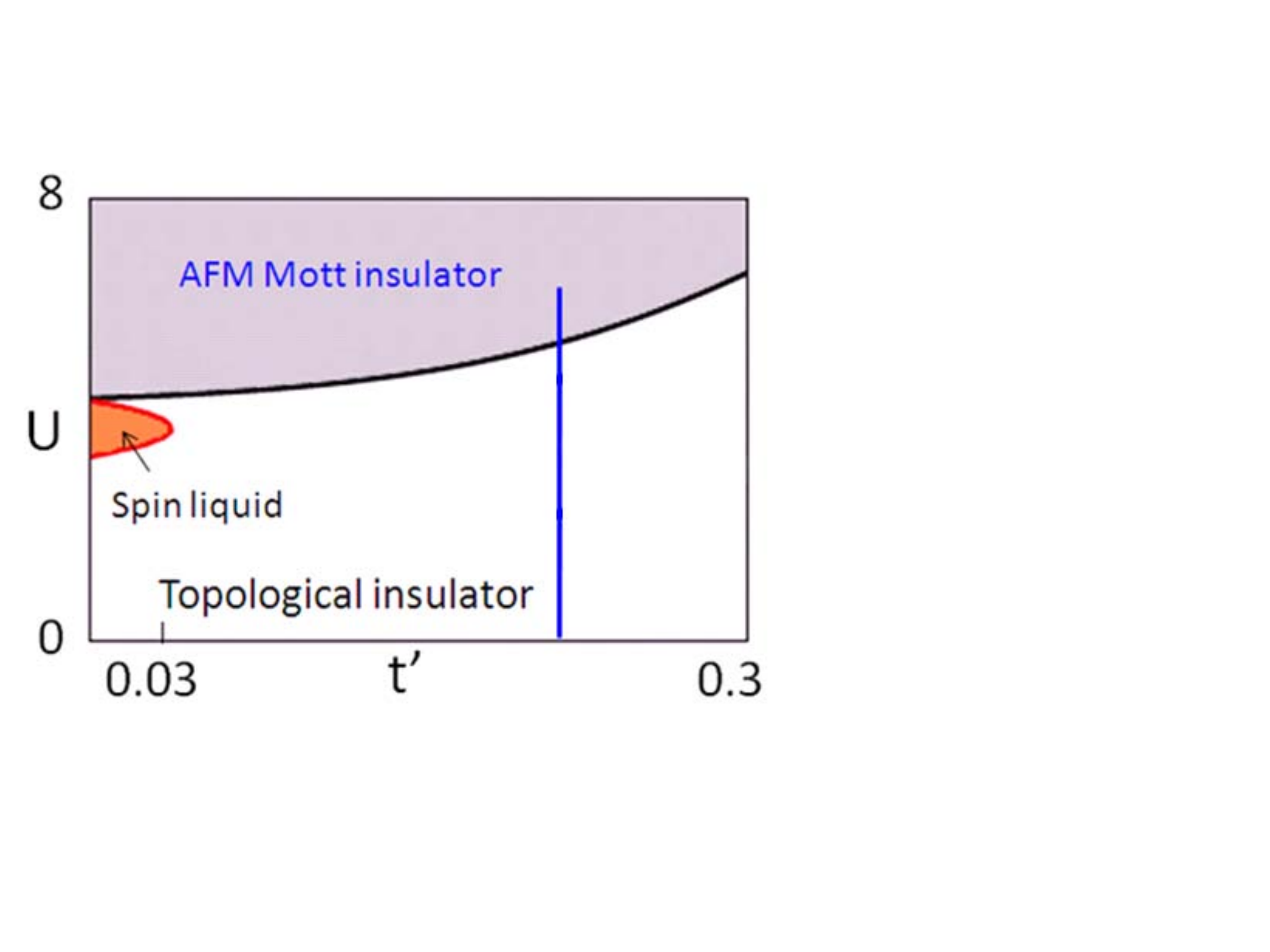}\vspace{-0.3 in}\caption{(color on-line) A schematic reproduction of the phase diagram of the $H_0+H_u$ reported in
Ref.\cite{assad}. The blue cut is considered in the text.\label{phdiag}}
\end{center}
\end{figure}

Recently two independent Monte-Carlo simulations\cite{assad,gm} were performed on the simplest kind of interacting QSHI. The Hamiltonian studied in these works is  $H_0+H_u$ where $H_0$ is the  $S_z$-conserving free electron model introduced by Kane and Mele\cite{km1}:
\be
H_0=\sum_{\s=\pm 1}\{-\sum_{\langle ij\rangle} c^+_{i\s}c_{j\s}+ i~t'\sum_{\langle\langle ij\rangle\rangle} \s~\nu_{ij} c^+_{i\s}c_{j\s}\}.\label{h0}\ee
 Here $i,j$ label the sites of a honeycomb lattice, the first term describes the nearest neighbor hopping, and the second term is a
spin dependent second neighbor hopping. Here $\nu_{ij}=(\hat{d}_1\times\hat{d}_2)_z/|(\hat{d}_1\times\hat{d}_2)_z|$ where $\hat{d}_1$ and $\hat{d}_2$ are unit vectors
along the two bonds the electron traverses when hopping from $j$
to $i$. $H_u$ is given by $H_u=U\sum_i n_{i\uparrow}n_{\downarrow}.$
According to Ref.\cite{assad}, for $t'\gtrsim 0.03$ (see \Fig{phdiag}) there are only two phases as a function of $U$. The large $U$ phase is an easy-plane (XY) antiferromagnetic (AF) Mott insulator; at small $U$ it is  an (interacting) QSHI with gapless spin and charge edge excitations.
\begin{figure}[tbp]
\begin{center}
\includegraphics[scale=0.4]
{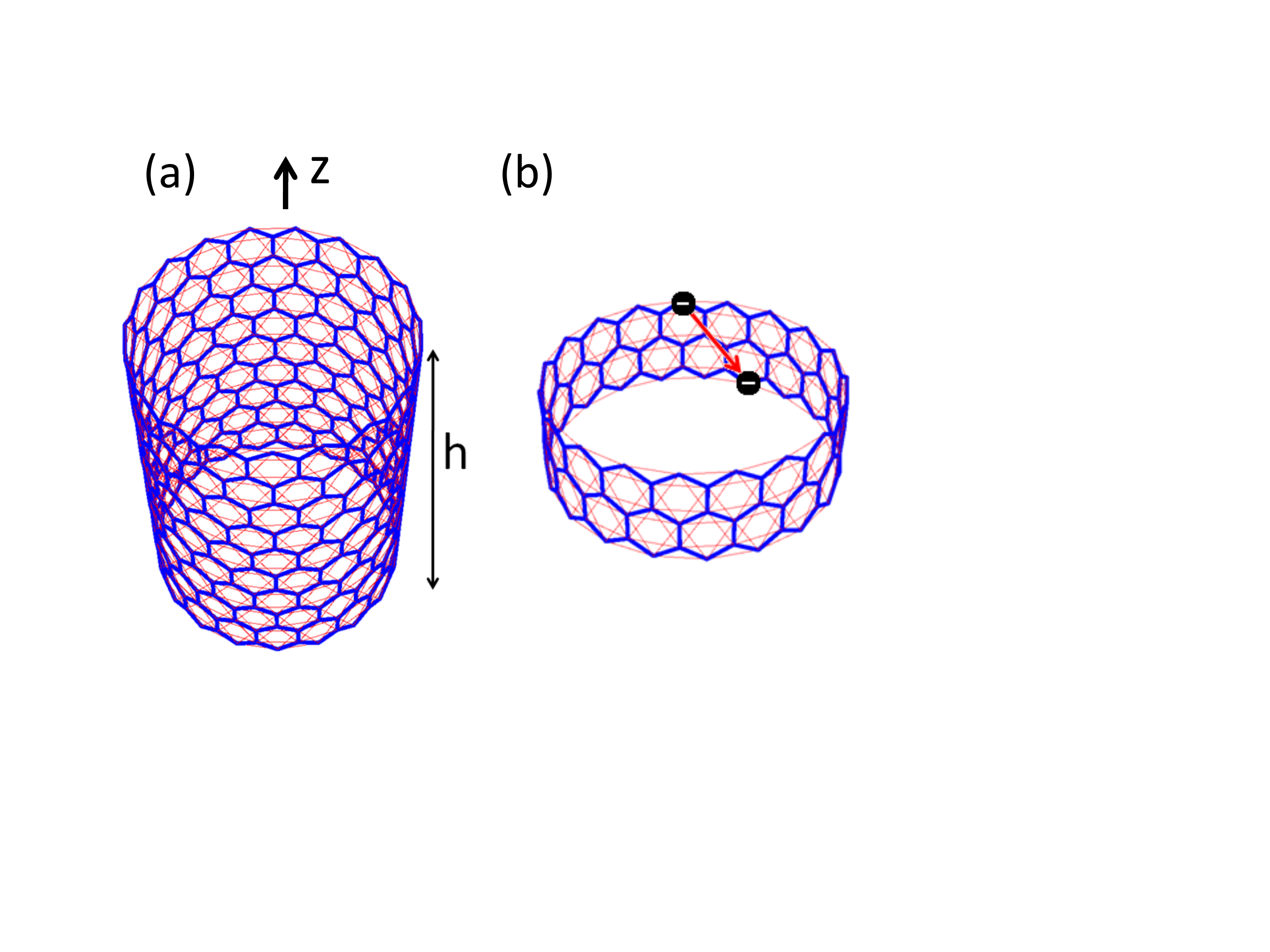}\vspace{-0.3in}\caption{(color on-line)(a) The spin-Hall insulator defined on a cylinder with height $h$. The blue/red bonds denote the nearest/second neighbor hopping. (b) When the cylinder is short, the electron can directly tunnel from one edge to the other.\label{cyl}}
\end{center}
\end{figure}

The present work is motivated by the following considerations.
Consider a system with edges (\Fig{cyl}(a)). At small $U$ the bulk is free of the magnetic moment and is a band insulator. At $U=0$ the low energy excitations are the helical edge modes described by the following Hamiltonian
\be
H_{E0}= \mp i v\int dx \Psi ^+\sigma _z\partial _x \Psi,\label{he}\ee
where $\mp$ applies to the top/bottom edges. In \Eq{he} $v$ is the edge velocity (which will be set to $1$ in the rest of the paper), and $\Psi$ is a two component fermion field whose first/second component corresponds to spin up/down. In mean-field theory the Hubbard U introduces a mass term - the  AF XY order parameter. Because the non-interacting theory has a logarithmically diverging susceptibility with respect to this order parameter, mean-field theory predicts AF XY ordered edges for arbitrarily small positive $U$. Of course, such order is destroyed by spin wave fluctuations; after that time reversal symmetry is restored. The resulting magnetic edges have power-law decaying AF XY correlation. Since these magnetic edge are quite different from the gapless edges in  the non-interacting theory, it is interesting  to ask in what sense is the resulting state a topological insulator.

\begin{figure}[tbp]
\begin{center}
\includegraphics[scale=0.4]
{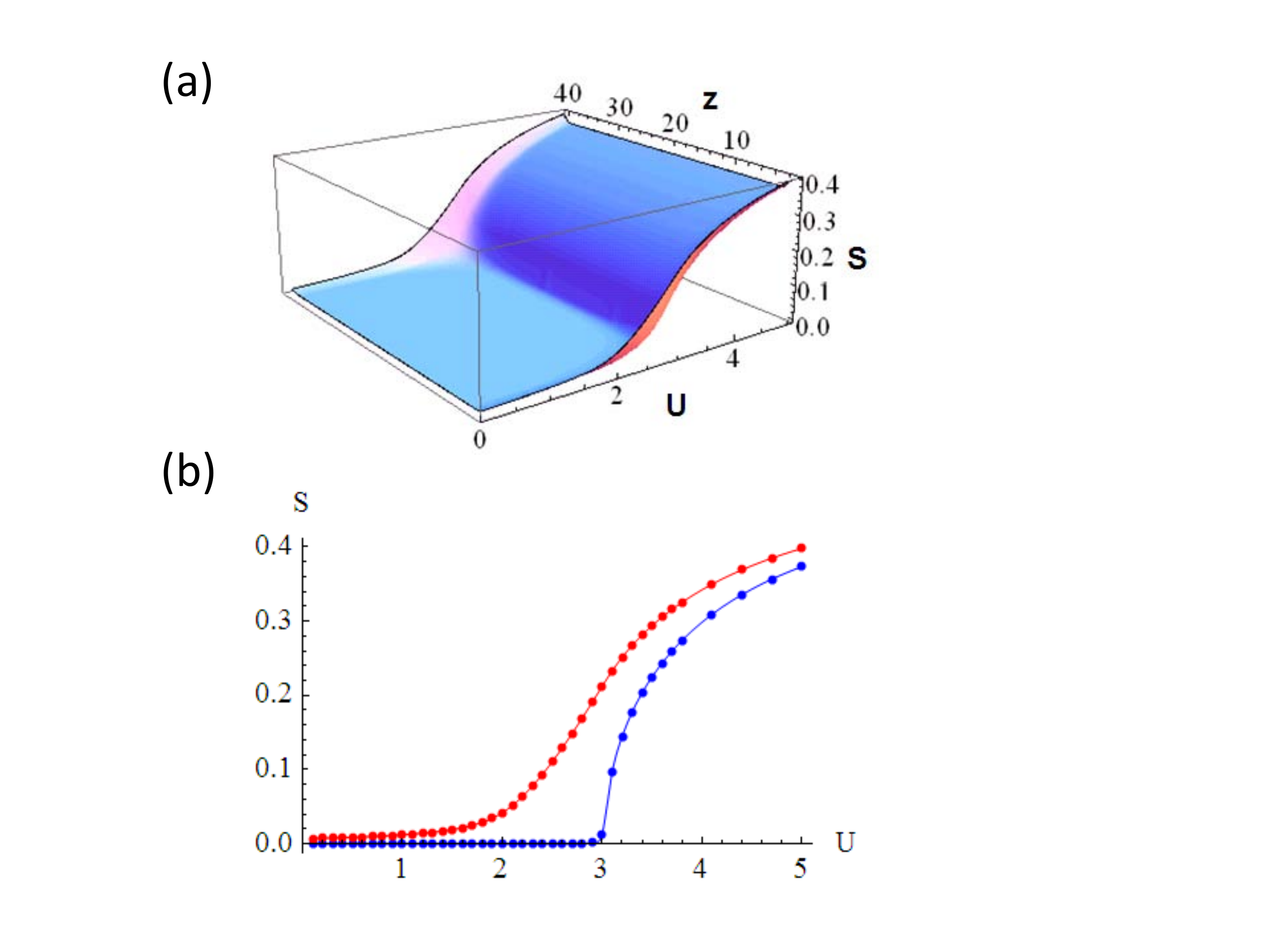}\caption{(color on-line) (a) The mean-field antiferromagnetic XY order parameter as a function of $z$ (see \Fig{cyl}(a) and $U$. The calculation is done for a cylinder with 35 unit cell in the periodic direction and 40 unit cells in the $z$ direction. The value $t'$ is 0.2. (b) The order parameter as a function of $U$ deep in the bulk $z=20$ (blue), and at the edge $z=1$ or $40$ (red). It is clear that while the order parameter in the bulk vanishes for $U\lesssim 3$ (the small rounding is due to finite size effect), the edge order parameter survives to the lowest $U$ value.\label{profile}}
\end{center}
\end{figure}

We start by studying the effects of Hubbard $U$ in mean-field theory.
Among all possible quadratic factorization, the following AF XY decoupling (the fact that AF is favored is dictated by
the edge wavefunctions)
\begin{widetext}
\be U \sum _in_{i\uparrow }n_{i\downarrow }\Rightarrow
-U\sum_i\left[\left\langle c_{i\uparrow }^+c_{i\downarrow }\right\rangle c_{i\downarrow }^+c_{i\uparrow }+\left\langle c_{i\downarrow }^+c_{i\uparrow }\right\rangle c_{i\uparrow }^+c_{i\downarrow }\right] + U\sum _i\left\langle c_{i\uparrow }^+c_{i\downarrow }\right\rangle \left\langle c_{i\downarrow }^+c_{i\uparrow }\right\rangle. \ee\end{widetext}
leads to a ``mass term'' of the free theory. 
In the presence of the XY order parameter the mean-field edge Hamiltonian read
\be
H_{E0}-m~[\cos\theta\int dx\Psi^\dagger\s_x\Psi
+\sin\theta\int dx\Psi^\dagger\s_y\Psi].\label{mf}\ee
 Since the free theory has a log-divergent susceptibility with respect to this order parameter, an infinitesimal positive $U$ induces the formation of the edge AF moments.
We check this prediction by performing a mean-field calculation on a finite cylinder for  $t'=0.2$ and $0\le U\le 5$ (this corresponds to the cut associated with the blue line interval in \Fig{phdiag}). The results are shown in \Fig{profile}; from which it is clear that while the order parameter deep in the bulk (the blue curve) vanishes for $U\lesssim 3$, the edge order parameter (the red curve) survives to the lowest $U$ value.
Thus for small $U$, mean-field theory predicts a one dimensional XY ordered antiferromagnetic at each (zigzag) edge. Spin waves destroy the long-range order and render the edge AF XY correlation power-law decaying. As a consequence time reversal symmetry is restored. However since the local moments introduces a single particle gap, one might wonder where are the gapless charge excitations.

In the presence of orientation fluctuations of the magnetic moments, $\theta$ in \Eq{mf} becomes position and time dependent. To describe that situation we need to replace \Eq{mf} by the following action
\begin{widetext}
\be
S_E=\int dx dt \{\bar{\Psi}\p_t\Psi\mp i \Psi ^+\sigma _z\partial _x \Psi - m \cos\theta(x ,t)~\Psi^\dagger \s_x\Psi - m\sin\theta(x,t)~\Psi^\dagger \s_y\Psi\}.\label{mf1}\ee\end{widetext}
Using the method of Abanov and Wiegmann\cite{aw} we have derived the stiffness term
\be
{K\over 2}\int dxdt~[(\p_t\theta)^2+(\p_x\theta)^2]\label{seff1}\ee by integrating out the fermions. The stiffness constant \be K={m^2\over 2\pi}\int_0^\Lambda {pdp\over (p^2+m^2)^2}={1\over 4\pi}(1-{m^2\over \Lambda^2+m^2}),\ee (here $\Lambda$ is a momentum cutoff) hence is less or equal to $1/4\pi$. The same result can be easily derived using bosonization in the $\Lambda\ra\infty$ limit. This implies the XY correlation function
\be
\langle e^{i\theta(0,0)}e^{-i\theta(x,t)}\rangle \sim (x^2+t^2)^{-{1/2\pi K}}\label{ss}\ee decays faster than
the Kosterlitz-Thouless bound $(x^2+t^2)^{-{1/4}}$\cite{KT}. Hence normally one would expect the power-law ordered AF edges to be unstable with respect to the proliferation of space-time vortices (vortex instantons). If so the edges will possess spin gaps.

In Ref.\cite{gw} Goldstone and Wilczek showed that after integrating out the gapped fermions in \Eq{mf1}, $\p_x\theta$ and $\p_t\theta$ gain coupling to the electromagnetic gauge field and the total effective action looks like
\be
S_{eff}=\int dxdt~\{{K\over 2}(\p_\mu\theta)^2-{i e\over 2 \pi}(A_0\p_x\theta-A_x\p_t\theta)\}.\label{seff}\ee
The last two terms of \Eq{seff} imply
the space and time gradients in $\theta$ produce  excess charge and current densities at the edges:
\be
\rho_E={e\over 2 \pi}\p_x\theta,~~J_E=-{e\over 2 \pi}\p_t\theta.\label{gwf}\ee Here $e$ is the electron charge.
 Because of \Eq{gwf} gapless spin wave excitations induce charge and current density fluctuations with the following correlation functions \be
&&\Pi_\rho(x,t)\propto \langle\p_x\theta(0,0)\p_x\theta(x,t)\rangle\sim {1\over x^2+ t^2}\nn
&&\Pi_J(x,t)\propto \langle\p_t\theta(0,0)\p_t\theta(x,t)\rangle\sim {1\over x^2+ t^2}.\ee
Thus gapless charge excitations emerge.

Because of \Eq{gwf}
a, e.g.,  a vorticity-$m$ instanton at the space-time location $(x_0,t_0)$ will cause
\be
&&\oint_{\p D} dx_\mu\p_\mu\theta =-{2\pi\over e}\oint_{\p D} dx_\mu \epsilon^{\mu\nu}J_{E,\nu}\nn&&=-{2\pi\over e}\int_D d^2x \p_\mu J_{E,\mu}=2\pi m.\label{loop}\ee Here $D$ is an arbitrary disk containing $(x_0,t_0)$ and
$J_{E,\mu}=(\rho_E,J_E)$ is the edge 2-current. \Eq{loop} implies
\be
\p_t\rho_E+\p_x J_E=-me~\delta(x-x_0)\delta(t-t_0),\ee  hence  vortex instantons violate the {\it edge} charge conservation. Nonetheless such instantons can occur through the the tunneling of electrons from one edge to the other\cite{note}(\Fig{cyl}(b)). Of course the amplitude of such tunneling is suppressed exponentially as a function of the cylinder height $h$. As a result the edge spin gap will be proportional to $e^{-\alpha h}$, which is not so different from the free electron case (only the $\alpha$ value is affected by the degree of relevance of the electron tunneling).

In the $h\ra \infty$ limit the (interacting) QSHI phase exhibits quantized  spin-Hall conductance. This can be understood as
follows. In the presence of an electric field between the two edges, a voltage difference $V$ develops. This induces a difference in $\p_x\theta$ between the two edges ($E_1$ and $E_2$) \be(\p_x\theta)_{E_1}-(\p_x\theta)_{E_2}={e V\over 2\pi K}.\ee  Because the spin current is ${K}\p_x\theta$, this gives \be J_{S_z}^{\rm {tot}}={e\over 2\pi}V\ee
hence the spin Hall conductance is ${e\over 2\pi}$ which is the same as the free electron value.
\begin{figure}[tbp]
\begin{center}
\includegraphics[scale=0.4]
{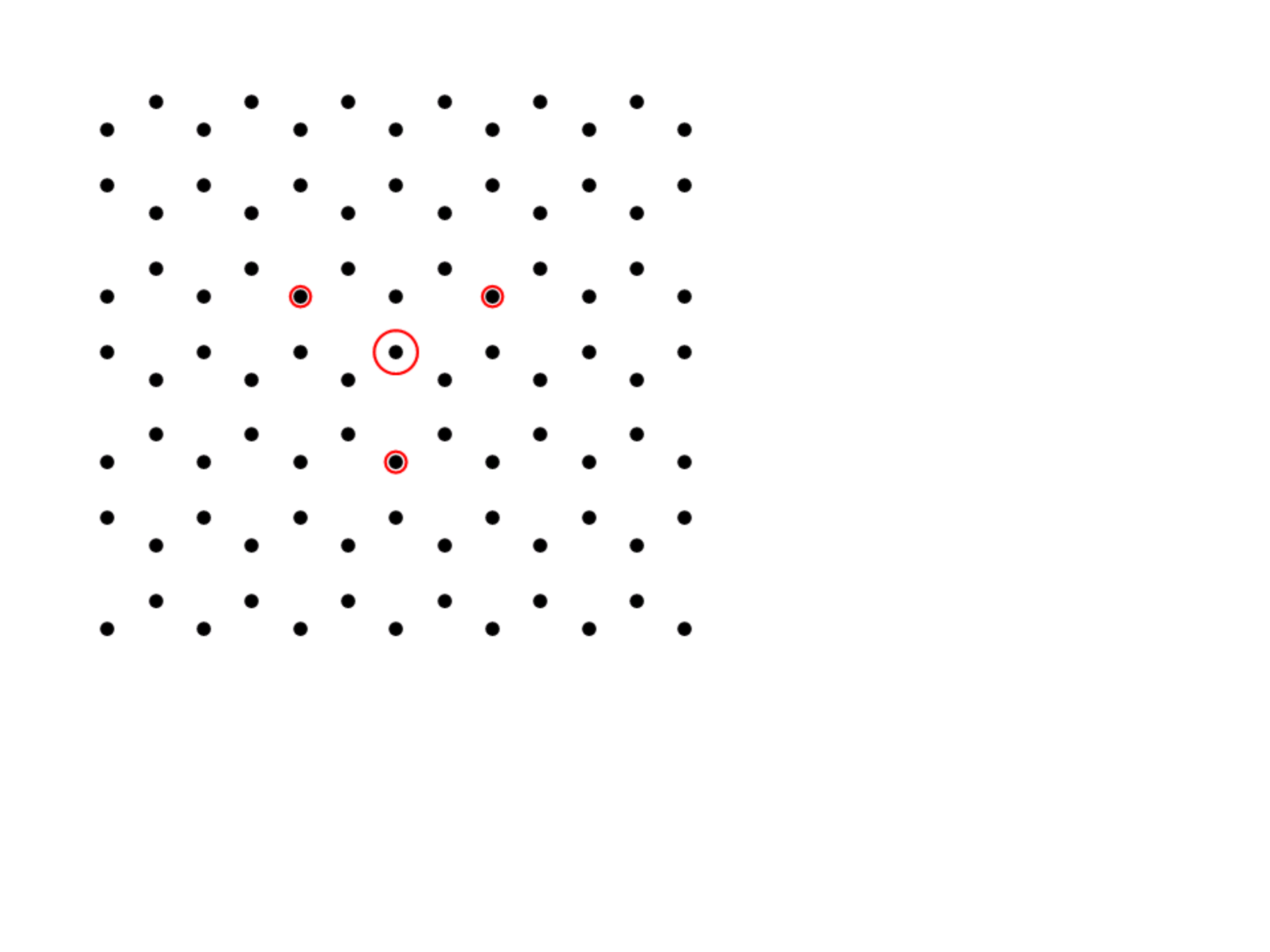}\vspace{-0.3in}\caption{(color on-line) An example of the XY order parameter profile associated with a localized magnetic exciton . The size of the red circles is proportional to the magnitude of the order parameter. \label{excit}}
\end{center}
\end{figure}


 Finally we consider the bulk transition between the AF Mott insulator and the QSHI. Let's starts from the QSHI phase. For this discussion let's use the periodic boundary condition. Let us consider, e.g., the blue cut in \Fig{phdiag}. First we approach the Mott insulator from the QSHI side. In the presence of $U$ its lowest-energy exciton is magnetic (we don't use the word ``triplet'' because $SU(2)$ is broken by the spin-orbit hopping down to $U(1)$). An example of the XY order parameter profile associated with a magnetic exciton is shown in \Fig{excit}. Now let's increase $U$. The transition into the AF insulator is triggered by the condensation of magnetic excitons. At the transition the modulus and the phase coherence of the AF XY magnetic order parameter form simultaneously.

It is also instructive to approach the transition from the AF Mott insulator side. In this case one naturally expects the XY order to be destroyed by the condensation of vortices. Because the AF XY order parameter, the triplet superconducting order parameter and the quantum spin Hall order parameter (which introduces the spin-dependent hopping term in \Eq{h0}) form a Wess-Zumino-Witten five-tuplet for the free-graphene bandstructure\cite{shinsei},
one expects the following charge density-skyrmion density relation\cite{aw}
\be
\rho={1\over 2\pi}\epsilon_{abc}n^a\p_x n^b\p_y n^c.\ee  Here $n^{1,2,3}$ is the unit vector
associated with the order parameter triad formed by the quantum spin Hall and the AF XY order parameters. The vortex charge is therefore proportional to $n_1 (1-n_1^2)$. Since $n_1=1$ (the modulus of the XY order parameter vanishes) at the Mott to quantum spin Hall insulator transition, hence the condensed vortices are charge neutral. As a result, the vortex condensed phase can be an insulator. In addition, since the modulus of the XY order parameter vanishes
at the transition the vortices do not see a background magnetic flux which frustrate the vortex condensation. This implies the universality class of the transition is three-dimensional XY like as claimed in Ref.\cite{assad}.\\

In summary, despite the apparent difference, the power-law correlated antiferromagnetic XY edges do exhibit properties expected for the quantum spin Hall insulators. The essential physics is the Goldstone-Wilczek mechanism; through which
the space-time gradients of the phase angle of the XY order parameter are proportional to the charge and current densities. The space-time vortices of the XY order parameter violate edge charge conservation hence are prohibited in thermodynamic samples. This is the mechanism through which the gapless charge and spin excitation are protected at the edges. At the moment we do not have a good picture for the ``spin liquid dome'' in \Fig{phdiag}. The main findings of this paper are summarized in the abstract.

{\it Acknowledgement:} I thank Tao Xiang, Guang-Ming Zhang, Shinsei Ryu and Hong Yao for helpful discussions. I am particularly grateful to Cenke Xu for explaining how to view the Mott to QSHI transition from the Mott insulating side to me. I acknowledge the support by the DOE grant number DE-AC02-05CH11231.


\begin{thebibliography}{99}
\bibitem{note} One might argue that an alternative tunneling process where an edge electron tunnel into the bulk and back can also allow vortex instantons. However in such virtual processes the intermdiate energy cost due to the bulk charge gap confine the instanton and anti-instanton.
\end{thebibliography}
\end{document}